\documentclass[prl,aps]{revtex4}

\usepackage{graphicx}
\usepackage{dcolumn}
\usepackage{bm}
\usepackage{epstopdf}

\begin{document}


\title{Super-Hard Superconductivity}

\author{D.P. Young, M. Moldovan, and P.W. Adams}
\affiliation{Department of Physics and Astronomy\\Louisiana State University\\Baton Rouge, Louisiana,
70803}%

\author{R. Prozorov}
\affiliation{Department of Physics and Astronomy\\University of South Carolina\\Columbia, SC  29208.}%

\date{\today}

\begin{abstract}
We present a study of the magnetic response of Type-II superconductivity in the extreme pinning limit, where screening currents within an order of magnitude of the Ginzburg-Landau depairing critical current density develop upon the application of a magnetic field.  We show that this "super-hard" limit is well approximated in highly disordered, cold drawn, Nb and V wires whose magnetization response is characterized by a cascade of Meissner-like phases, each terminated by a catastrophic collapse of the magnetization.   Direct magneto-optic measurements of the flux penetration depth in the virgin magnetization branch are in excellent agreement with the exponential model in which $J_c(B)=J_{co}\exp(-B/B_o)$, where $J_{co}\sim5\rm{x}10^6$ A/cm$^2$ for Nb.  The implications for the fundamental limiting hardness of a superconductor are discussed. 
\end{abstract}

\maketitle

	In contrast to Type-I superconductors, Type-II systems support a mixed state consisting of an array of Abrikosov vortex lines \cite{Tinkham,Abrikosov}.  Technologies requiring high current densities and/or high fields typically utilize ÔhardÕ Type-II superconductors in which dissipative vortex flow is inhibited by deep vortex pinning centers \cite{Wilson,Poole}.  Here we explore the fundamental limiting hardness of a superconductor by addressing the thermodynamic, magnetic, and macroscopic quantum characteristics of superconductivity in the presence of pinning forces of sufficient strength to support current densities approaching the Ginzburg-Landau depairing critical current, $J_{c}^{GL}$ \cite{Tinkham}.  In this limit, local spontaneous vortex nucleation becomes energetically possible due to the suppression of the superconducting energy gap.  Furthermore, stretching and twisting of the vortex lines can lead to the release of segment loops \cite{Schwarz,Poole}.  Both of these processes will relax flux gradients and thereby produce dissipation.  Using magnetization, magneto-thermal, and magneto-optical probes, we show that this Òmaximum pinning limitÓ is well approximated in cold drawn Nb and V wires, where screening currents of order 10 - 25\% of $J_{c}^{GL}$  develop upon the application of a magnetic field.  Their magnetic response appears to be a convolution of Type-I and Type-II behaviour, with cascading Meissner-like phases, each terminated at a magnetization energy that exceeds the thermodynamic limits of the post-annealed samples \cite{Tinkham}.  In addition, we show that pinning in these materials is purely a morphological effect which can be tuned from the maximum pinning limit to the weak pinning limit via annealing.

	The response of a superconductor to an applied magnetic field in the presence of strong pinning is determined in large degree by the competition between vortex pinning forces and the Lorenz force arising from induced screening currents.  In the early 1960's Bean \cite{Bean} introduced a phenomenological model in which the application of a magnetic field $H$ produces a non-equilibrium flux gradient determined by a depinning critical current density $J_c$.   The model assumes that after an initial transient, a critical state is reached where $J_c$ flows in the regions of flux penetration.   Bean also assumed that $J_c$ was independent of field, but later extensions of the model included field dependences \cite{Poole}.  Perhaps the most successful of the enhanced Bean-type models is the exponential model \cite{ExpModel} in which,
 \begin{equation}
 J_c=J_{co}\exp(-B/B_o),
 \end{equation}
where $B_o$ is a phenomenological characteristic field, and $J_{co}$ is the intrinsic critical current density.  The model is most transparent when applied to the virgin magnetization response of a zero-field-cooled superconductor.  Solving Maxwell's equation $\mu_o{\bf J}={\bf \nabla} \rm{x} {\bf B}$ in a cylindrical geometry of radius R and length L$\gg$R, one finds that upon the application of an external field $H>H_{c1}$ vortices enter from the surface and penetrate to a depth $d$.   
\begin{equation}
d=d_o[\exp(\mu_oH/B_o)-1],
\end{equation}
where $d_o=B_o/(\mu_oJ_{co})$ is the characteristic penetration depth.   Gradients in the vortex density produce Bean currents that flow within this annulus and serve to shield the applied field.  If $d<R$ then the induction interior to $d$ is zero ($B=0$).  At sufficiently high  fields, however, flux will penetrate to the center, $d\geq R$.  In either case, the resulting magnetization is simply the volume average $<B(\xi)/\mu_o-H>$ where $\xi = R -r$ \cite{Bean}, and $B(\xi)$ satisfies the boundary conditions $B(0)=\mu_oH$ and $B(d)=0$ for $d<R$,
\begin{eqnarray}
B(\xi) = & B_o\ln[(\gamma d_o-\xi)/d_o] &\; \; \; \xi<d \nonumber \\  
 = & 0 & \; \; \; \xi\geq d,
\end{eqnarray}
where $\gamma=\exp(\mu_oH/B_o)$.  

	Though Bean-type models have proven to be quite useful in the analysis of the magnetization curves of superconductors with moderate to strong pinning, it is well known that the low temperature magnetization response of extremely hard superconductors is often discontinuous, thus limiting the applicability of the models to those systems \cite{FluxJumps}.  In the context of the above exponential model, the "hardness" is in large part reflected in the magnitude of the intrinsic critical current density $J_{co}$.  One can obtain an estimate of the maximum possible magnitude of this current density, $J_{co}^{max}$ by considering the pinning force associated with a vortex of length L whose core, of radius $\xi$, sits in a rectilinear void of length L and radius $\xi$ \cite{Brandt1}.  By placing its core in the void, the vortex line gains an energy per unit length $U={B_c^2\pi\xi^2}/{2\mu_o}$ with a corresponding pinning force $F_p\approx U/2\xi$.  Equating this pinning force with the Lorentz force $F_L=\Phi_oJ_{co}^{max}$ one obtains \cite{Kunchur},
\begin{equation}
J_{co}^{max}=\frac{3\sqrt{3}}{32}J_c^{GL}\sim 0.16J_c^{GL}. 
\end{equation}
 In this limit maximum hardness limit an applied field can produce extremely high magnetization energies, which in turn lead to well documented thermomagnetic collapses of the magnetization \cite{Wipf}.  Indeed, more than 40 years ago, Kim and co-workers \cite{Kim} reported discontinuous $M$-$H$ curves in thin-wall tubes of sintered Nb$_3$Sn powders in which the magnetization energies obtained from their ÒhardestÓ samples were well above the thermodynamic critical field of Nb$_3$Sn.  Unfortunately, little could be learned from modeling the  magnetization curves, since they were Meissner-like \cite{Brandt2}; $M\approx-H$ up to the point of collapse.    In the present Letter, we use magneto-optical (MO) techniques to directly measure the flux penetration depth in the zero-field-cooled state of cold drawn Nb wire in order to extract the two independent parameters of the model, $J_{co}$ and $B_o$, independent of the magnetization measurements. 
 		 
	Magnetization measurements were made on 1-mm diameter Nb (99.8 \%) and V (99.5\%) wire segments obtained from Alpha Aesar.  The magnetization was measured on a Quantum Design PPMS system with a base temperature of 1.8 K and maximum field of 9 T.  The magnetization samples consisted of wire segments of length $\sim6$ mm that were sandblasted both to remove the patina and to ensure a uniformly disordered surface.  Magneto-optical (MO) measurements \cite{MO} with a spatial resolution of  $\sim5$ $\mu$m were performed on 0.5-mm thick disks cut and polished from the wire stock.  The un-annealed Nb wire, from which most of the data were taken, had a 10-K resistivity of ~1 $\mu\Omega$-cm and a corresponding mean free path $l_o\sim30$ nm.  The magnetization and MO measurements were repeated after vacuum annealing the samples at 1100 $^o$C for 24 hrs. 
	
	Shown in Fig.\ 1 are the low temperature magnetization curves of annealed and un-annealed Nb and V wire segments, where the magnetic field was applied along the cylindrical axis.  Nb ($T_c = 9.2$ K) and V ($T_c = 5.4$ K) are two of only three known Type-II elemental superconductors \cite{Poole}, and indeed, the annealed magnetization curves (open symbols) represent classic Type-II behaviour.   The un-annealed curves, however, are profoundly different both in structure and scale and would not be recognizable to most condensed matter physicists as associated with Type-II superconductivity.  Several salient features are worth noting. First, the magnetization energies are extremely high.  We have extracted the thermodynamic critical fields from the annealed data by integrating the average magnetization over one half of a field cycle and find that  $H_c = 153$ mT and 51 mT for Nb and V, respectively. The magnetization of the zero-field-cooled branch, labelled ZFC in Fig.\ 1, reaches a maximum value $M_{cr}\sim -2H_c$ in the Nb sample and $M_{cr}\sim-4H_c$ in the V sample, which are somewhat higher than what is typically observed in HTC systems were $M_{cr}\sim H_c$.   Second, upon reaching $M_{cr }$ the Nb magnetization catastrophically collapses to a value that crosses the annealed curve. By attaching miniature thermocouples to the samples we were able to verify that the Nb wires, in particular, were heated above $T_c$ by the collapse, therefore the re-equilibration occurred along a field-cooled path.   As $H$ is increased beyond the first collapse, a series of Meissner-like branches develops
 \begin{equation}
	M^{n+1}=-(H-H_{cr}^{n}),
\end{equation}
where  $H_{cr}^{n}$ is the critical field of the $n$Õth collapse. Finally, one might expect that the largest $M_{cr}$ would be observed on the ZFC branch, but in fact, the largest magnetization magnitude develops on the zero-crossing-branch (ZCB) in Fig.\ 1.

	There is little doubt that micro-structural defects induced by the drawing process are producing extremely strong pinning centres in the Nb and V wires, and that the pinning is allowing anomalously high magnetization currents to develop.  In Fig.\ 2 we show scanning electron micrographs of an un-annealed and annealed Nb wire after chemical etching (see Fig.\ 2 caption).  Etching produced no significant changes in the $M$-$H$ behaviour of the un-annealed sample, and the micrographs clearly indicate that structural disorder on the scale of microns exists throughout the volume of the un-annealed wires.  In contrast, the annealed samples exhibit crystalline grain boundaries as indicated by the arrows.  

	The magnetization collapses in Fig.\ 1 occur due to the fact that we cross the thermomagnetic stability field, $B_{sf}$.  A rough estimate of the ZFC stability field of Nb at $T=3$ K is given by $B_{sf}\approx\sqrt{\mu_oc_v(T_c-T)}\sim 250\;\rm{mT}$ which agrees well with the ZFC branch in Fig.\ 1, where $c_v\sim8\rm{x}10^3$ J/m$^3$K is the average volume specific heat of Nb \cite{NbSH} below $T_c$.  Clearly, however, the ZFC branch deviates little from $M=-H$ up to the first collapse making it difficult to accurately extract $J_{co}$ and $B_o$ from a fit to the magnetization data or to even establish the appropriateness of the exponential model for this system.  To address this issue we employed a standard MO technique to get a direct image of the depth and magnitude of flux penetration into both the un-annealed and annealed samples.  In Fig.\ 3 we present MO images of a Nb disk of length 0.5 mm cut from the 1-mm diameter Nb wire stock.  The magnetic field was oriented along the disk axis.  Note that after ramping to the maximum applied field of the MO system, 135 mT, only a thin shell of vorticity enters the un-annealed disk, in stark contrast to the extensive flux penetration in the annealed one.  This is most clearly seen in the field profile plots in Fig.\ 4.  

	In Fig.\ 5 we plot the penetration depth, $d$, in the ZFC branch of the unannealed Nb disk as a function of $H$, using the criterion that $B/\mu_oH=0.5$ in the profile data of Fig.\ 4.  Note that $d$ is not simply proportional to $H$.  The solid line is a least-squares fit to Eq.(2) in which $d_o$ and $B_o$ were independently varied \cite{Demag}.  Not only is the overall quality of the fit in Fig.\ 5 very good, but the extracted values of $J_{co}=5.8\rm{x}10^6$ A/cm$^2$ and $B_o=62$ mT agree very well with values obtain from fits to the ZFC magnetization using Eqs.\ (2) and (3).  We note that $J_{co}$ is a factor of 4 higher than reported for hardest Nb$_3$Sn tubes in Ref.\ 7 and an order of magnitude higher the critical currents of cold worked Nb-25\%Zr wires of Ref.\ 11.  Indeed, recent measurements of the transport critical current density in Nb films \cite{NbCC} report a depairing critical current $J_{c}^{GL}\approx5\rm{x}10^7$ A/cm$^{2}$   at $T$ = 3 K, indicating that $J_{co}\sim0.15 J_{c}^{GL}$ and $d_o \sim15 \lambda$.  Interestingly, these critical currents are quite close to the maximum depinning current of Eq.\ (4), lending further evidence that the unannealed Nb wire is near its maximum hardness.  Similar strong pinning effects have recently been reported in twin boundaries of YBa$_2$Cu$_3$O$_{7-\delta}$ \cite{SPYBCO} and at the edge barriers of MoGe films \cite{SPMoGe}.  Though neither of these systems exhibit significant bulk pinning, they provide compelling evidence that it is possible for pinning potentials to be of sufficient strength so as to accommodate depairing limited screening currents.    

	 The ZFC magnetization collapses at $\mu_oH_{cr}^{1}\sim300$ mT which from Eq.\ (2) corresponds to $d\approx 0.25\;\rm{mm}\approx R/4$.  Unfortunately, we are not able to extend the MO measurements beyond 135 mT, and therefore do not have penetration depth measurements on the subsequent branches nor up to the first collapse.   It seems likely that both $J_{co}$ and $B_o$ are significantly suppressed in the higher field branches.  Indeed, we find the magnitude of the critical magnetization $M_{cr}^n$ in the Nb sample is proportional to $[1-(H_{ave}/H_{c2})^2]$, where $H_{ave}=(H_{cr}^{n}+H_{cr}^{n-1})/2$ is the average field within the penetration depth $d$.  This observation is consistent with the largest magnetization developing in the zero crossing branch (ZCB) in Fig.\ 1.  Extending the MO measurements across the entire field range of Fig.\ 1 should prove invaluable in modeling the higher field branches and, in particular,  the behavior near $H_{c2}$.
	 
	In summary, we have correlated magneto-optical measurements of the flux penetration depth with the magnetization behavior of extremely hard superconducting Nb wires.  We find that the virgin magnetic response is very well described by the exponential model with intrinsic critical current densities within an order of magnitude of the depairing critical current.  Our analysis suggests that the exponential field dependence of the critical current is, in fact, a fundamental property of the condensate in this extreme pinning limit.  Further studies of the topological characteristics of the pinning centers along with the metallugical aspects of the drawing process could provide insights into pinning processes in general.  Furthermore, the ability to systematically decrease the pinning strength via annealing treatments affords one the opportunity to study the evolution of the system from the super-hard limit to the weak pinning limit and perhaps gain a clearer understanding of the pinning limits of other more technologically important materials.
	 
	 We gratefully acknowledge enlightening discussions with Dana Browne, Milind Kunchur, Ernst Brandt, and Ilya Vekhter.  This work was supported by the National Science Foundation under Grant DMR 02-04871.


\newpage

\begin{figure}
\includegraphics[width=5in]{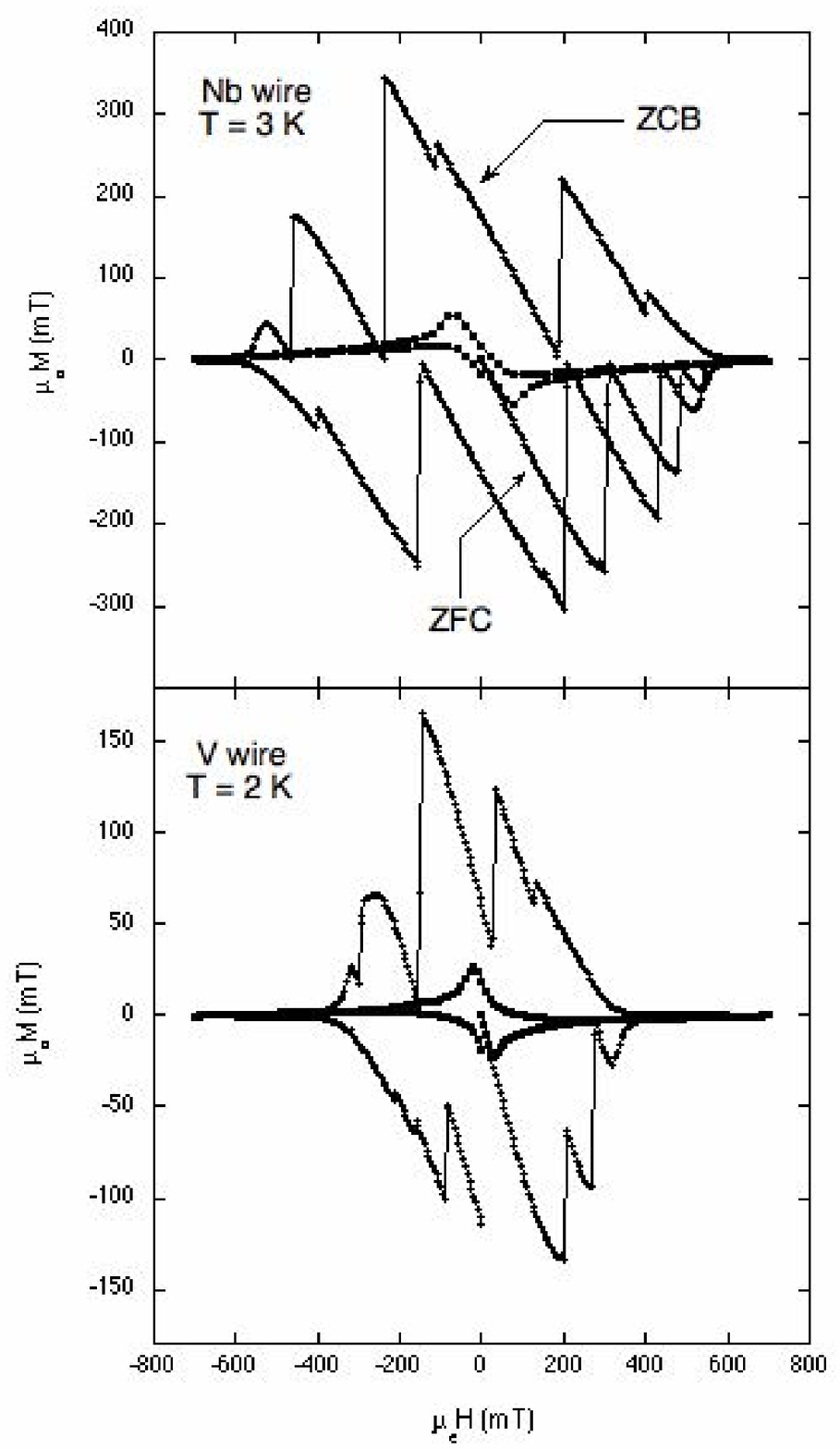}
\caption{\label{M-H} Magnetization of annealed (open symbols) and un-annealed (filled symbols) Nb and V wires as a function of field applied along the wire axis.  From a standard analysis of the annealed data we obtain $\mu_oH_{c1} = 76$ mT, $\mu_oH_{c2}= 750$ mT, $\mu_oH_{c} = 153$ mT, $\lambda =30$ nm, and $\kappa = 1.6$ for Nb, and $\mu_oH_{c1} = 25$ mT, $\mu_oH_{c2}= 680$ mT, $\mu_oH_{c} = 51$ mT, $\lambda =100$ nm, and $\kappa = 4.5$ for V. }
\newpage
\end{figure}

\begin{figure}
\includegraphics[width=5in]{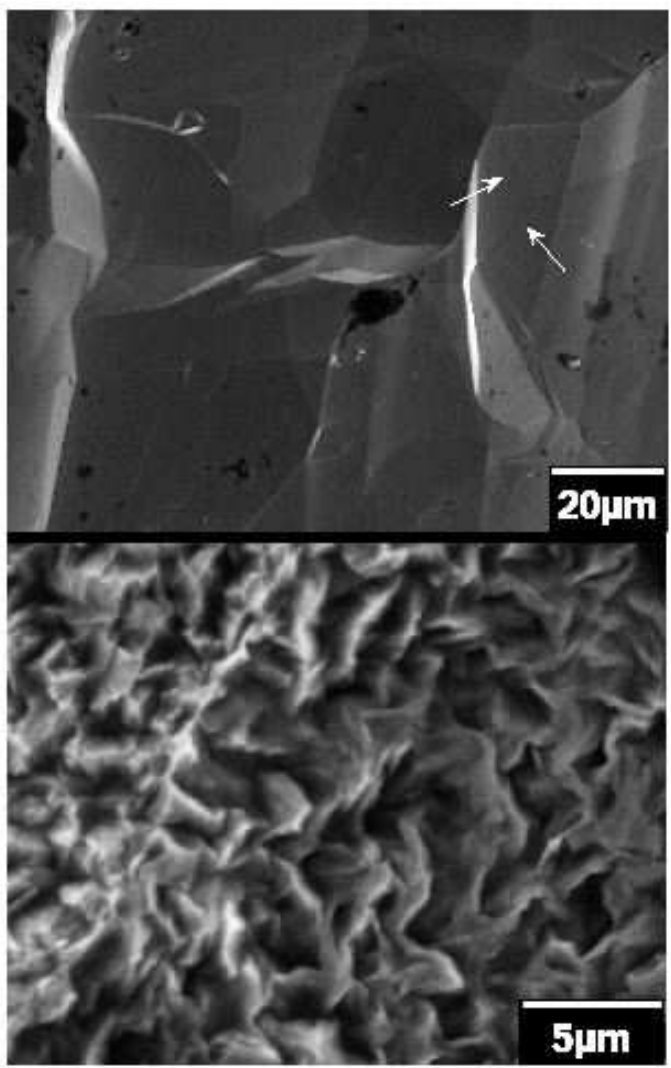}
\caption{\label{SEM}  (A) Scanning electron micrographs of the surface of Nb wire after vacuum annealing at 1100 $^o$C for 24 hrs.  The arrows indicate crystalline grain boundaries. (B) Surface of the un-annealed wire. The surfaces were prepared by a 5 min etch in a 1:1:2 solution of HNO$_3$, HF, and HSO$_4$.}
\newpage
\end{figure}

\begin{figure}
\includegraphics[width=5in]{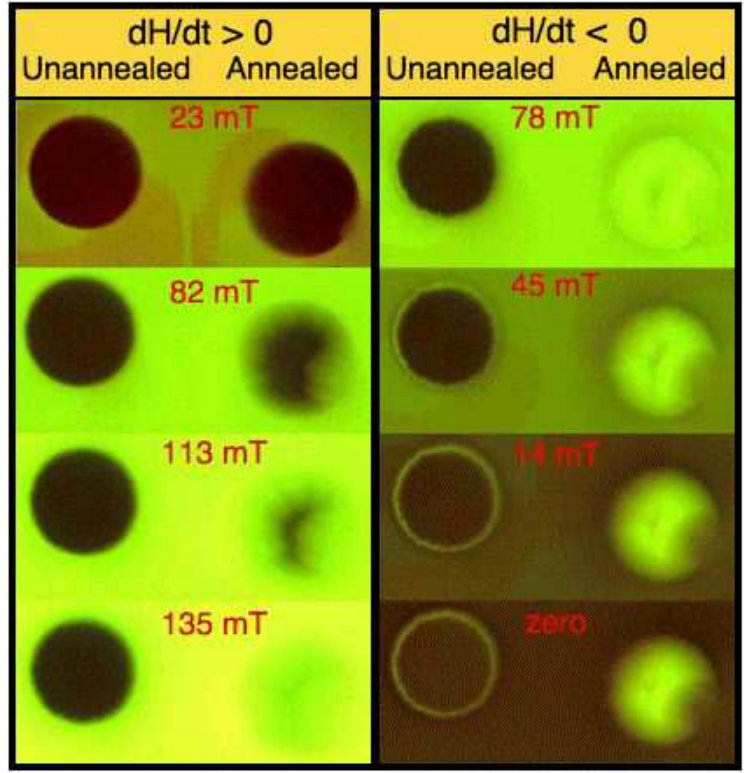}
\caption{\label{MO} Magneto-optical images of an annealed and un-annealed Nb disk at T = 3 K with field oriented along the disk axis.  The left column corresponds to increasing field and the right to decreasing field.  Dark corresponds to zero field.}
\newpage
\end{figure}

\begin{figure}
\includegraphics[width=5in]{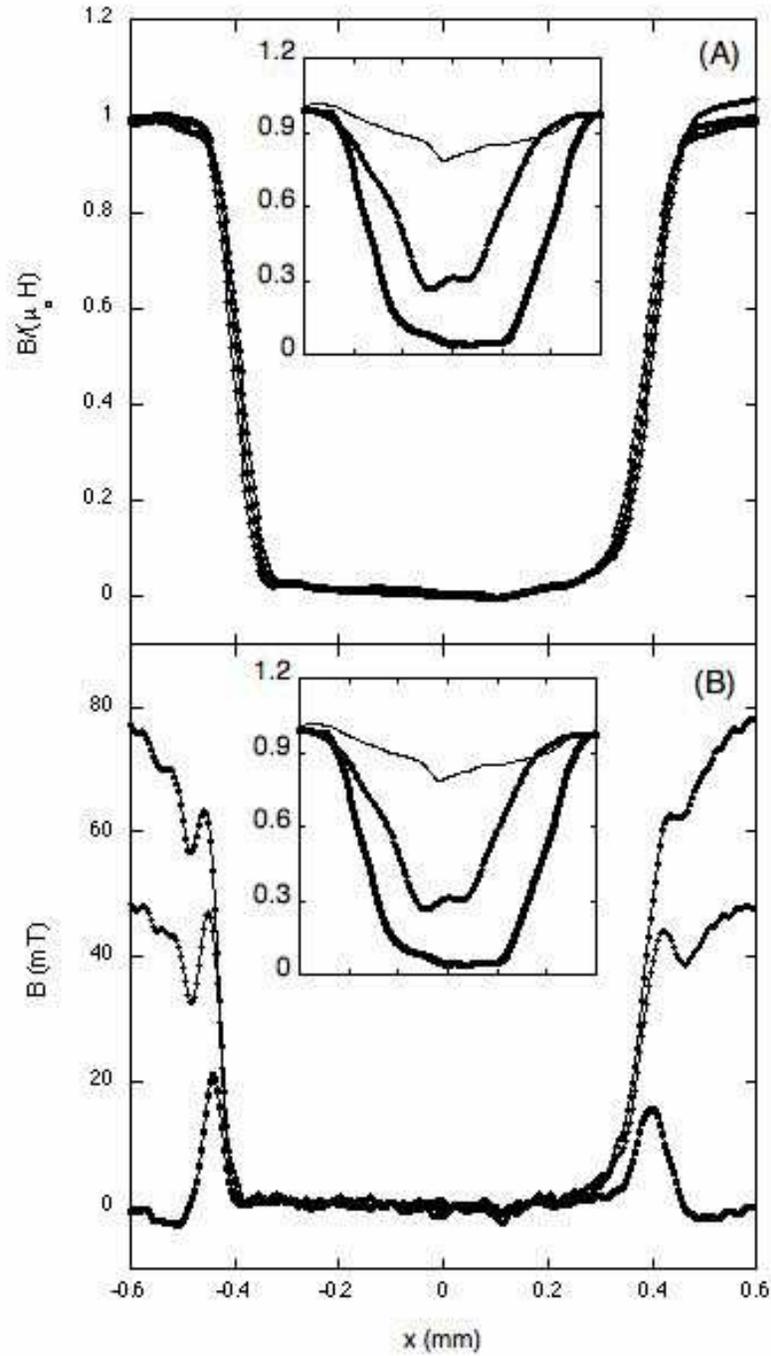}
\caption{\label{Profiles} Field profiles obtained from the MO images in Fig. 3.  (A) Increasing field profiles normalized by the applied field, $\mu_oH = $23, 82, 113, and 135 mT.   (B) Decreasing field profiles after reaching a maximum field of 135 mT. Panel A and (B) insets: profiles of the annealed sample at increasing fields of $\mu_oH = $14, 45, and 78 mT and decreasing fields of $\mu_oH =$ 75, 45, and 0 mT. }
\newpage
\end{figure}

\begin{figure}
\includegraphics[width=5in]{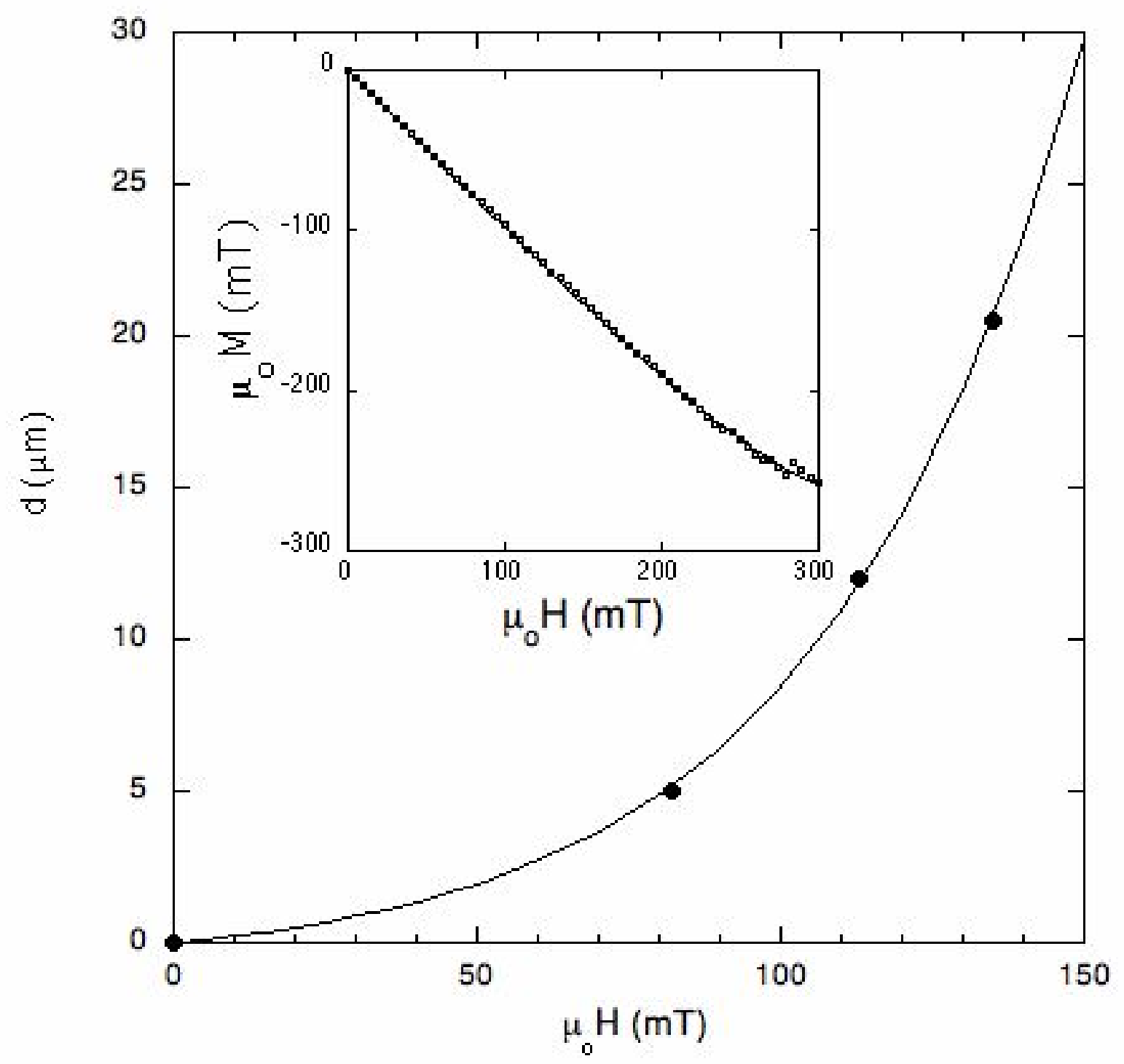}
\caption{\label{d-H}  Penetration depth as a function of applied field as determined from the data in Fig.\ 4.  The solid line is a best fit to Eq.(2) with  $J_{co}=5.8\rm{x}10^6$ A/cm$^2$ and $B_o=62$ mT independently varied.  Inset: ZFC branch of the Nb magnetization curve in Fig.\ 1.  The solid line is a fit to the volume average $<(B(\xi)-\mu_oH)>$, where $B(\xi)$ is given by Eq.(3) and $J_{co}=5.2\rm{x}10^6$ A/cm$^2$ and $B_o=92$ mT were varied.}
\newpage
\end{figure}



\begin{references}
\bibitem{Demag} The demagnetization factor of the disk, $N=1/3$, was taken into account in the determination of $B_o$.
\end{references}
\end{document}